\documentstyle[multicol,aps]{revtex}

\title{Corrections to scaling in systems with thermodynamic constraints}

\author{I. M. Mryglod$^{1,2}$ and R. Folk$^2$}

\address{$^1$Institute for Condensed Matter Physics,
         1 Svientsitsky Street, UA-79011 Lviv, Ukraine}
\address{$^2$Institute for Theoretical Physics, Linz University,
         A-4040 Linz, Austria}

\date{\today}

\begin{document}

\maketitle

\begin{abstract}

Using thermodynamic arguments treatment it is shown that, independetly on
whether Fisher renormalization changes the critical exponents near a phase
transition in a constrained system or not, new corrections to scaling with correction
exponents proportional to the specific heat index $\alpha$ appear. Because of
the smallness $\alpha$ for the Ising, the XY, and the Heisenberg universality
classes these corrections are dominant and can cause strong crossover effects.
It is proven that the appearance of Fisher corrections to scaling is a quite
general feature of the systems with constraints.

\vspace{12pt}
\noindent
Pacs numbers: 05.70.Fh; 64.70.-i; 75.50.-y; 75.40.-s

\vspace{12pt}

\end{abstract}

\begin{multicols}{2}

\newpage
\vspace{1cm}
The theoretical study of phase transitions in classical spin models has played
an important role in the understanding of the key features of critical behaviour
in real systems. It is well known that each model like the classical Ising, the
XY and the Heisenberg model represents a certain universality class with a
specific set of critical exponents. Apart from the spatial dimensionality and
the range of the interactions  a universality class is characterized by the
number of components of the order parameter.  In contrast to the case of ideal
spin lattice models, real systems are exposed to various kinds of imperfections
(lattice defects, impurities, etc.) which can have a significant influence on
the true critical behaviour of the system considered. In many practical
applications imperfections in the real systems result in a certain thermodynamic
constraint. In the recent years, of notable interest were continuum spin fluid
models \cite{Hem,Kal,Kob,Vak2} which, in particular, are considered as a first
step towards the modelling of ferrofluids \cite{Feld} and adsorption surface
phenomena \cite{Sme}. Several important results, which have both theoretical
and experimental interest, were obtained for such models. For example, it
was found that, because of the interplay between spin and translational degrees
of freedom, the phase diagrams for spin fluids are much more complicate
\cite{Hem,Stel1,Stel2,Fei,Lom,Tav1,Tav2} compared with pure liquids and could
lead to the magnetic ordered state both in the gaseous and liquid phases. By
applying an external magnetic field one could also shift significantly the locus
of a gas-liquid transition \cite{Folk,Lado,Sokol} and change the dynamic
properties \cite{Mryglod,MryglodFolk,FolkMoz}; both static and dynamic properties
in the spin fluid models show the differences from pure fluids and/or spin lattice
models. Note that in the case of spin liquid models we deal, in fact, with a
system with thermodynamic constraint, namely, fixing the density of particles.
Therefore, the general question arises how a thermodynamic constraint affects
the critical singularities of the system considered. The answer to this question
has been partly given a long time ago by M.Fisher \cite{Fisher}. In particular,
it was shown that, due to the constraint, the main singularities in the
constrained system are renormalized if the critical exponent of the specific heat
$\alpha$ is positive or remain the same as in the `ideal system' (without
constraint) when $\alpha$ is negative. This effect is known in the literature as
`Fisher renormalization', and for many examples of model and real systems with
constraints this renormalization of critical exponents has been proven.

In Ref.~\cite{Nijmei1} it was found by Monte Carlo simulations that the critical
exponents of a three-dimensional Heisenberg fluid model are in disagreement with
the results known previously for the lattice model. Similar findings were later
obtained for three- and two-dimensional (3d and 2d) Ising spin fluids
\cite{Nijmei2,Ferreira}. We note that the main difference between the cases of
3d Ising and 3d Heisenberg fluids, as systems with an annealed disorder, is that
according to the Fisher renormalization \cite{Fisher} the critical exponents
have to be renormalized in the first case and remain the same as in the pure
system in the second one. However, in both cases systematic deviations from the
critical exponents predicted by theory were found in Monte Carlo calculations. We
note also that in Refs.\cite{Nijmei1,Nijmei2} a weak dependence of universal
quantities on a density of particles $n=N/V$ was observed. From these facts we
conclude that the critical behaviour in such fluids with internal degrees of
freedom are strongly influenced by nonlinear crossover effects which hide the
true asymptotic critical behaviour leading to effective critical exponents
in power law fits over the restricted temperature regions considered.

In this paper we will focus on the study of continuous spin fluid models and the
main problem to be discussed is whether the magnetic transition in a fluid spin
model lies in the same universality class as the corresponding transition in the
lattice model and, if so, what is the reason for such a strong crossover, observed
for these systems in Monte Carlo simulations \cite{Nijmei1,Nijmei2,Ferreira}. In
this paper we will follow the scheme of thermodynamic argumentation developed in
the original paper of Fisher \cite{Fisher} more than thirty years ago. Due to the
general arguments used, we believe that our results are of interest also to many
other systems with thermodynamic constraints.

\section{General background}

Let us consider the grand canonical ensemble with the thermodynamic potential
$\Omega = \Omega [\mu,h,T]$ of a continuum spin liquid at fixed volume $V$.
The pressure of the system is simply connected with $\Omega [\mu,h,T]$, namely,
$P= - \Omega/V$, and from the thermodynamics we have the well-known expression
\begin{equation} \label{Rel1}
dP = n d\mu + m dh + s dT,
\end{equation}
where $n$, $m$, and $s$ are the particles' density, the magnetization and
entrophy per unity of the volume, respectively. The conjugated thermodynamic
fields in (\ref{Rel1}) are the chemical potential $\mu$, the external
magnetic field $h$ and the temperature $T$, respectively. In this study we
are interested in the ferromagnetic phase transition of second order with $m$
being the order parameter.

The first (and the most important) assumption, which has been made by Fisher
in Ref.~\cite{Fisher} and will be also used in our treatment, is that {\it the
nature of the phase transition remains ideal if observed at fixed  `hidden'
thermodynamic force} which is directly related in our case to chemical potential
$\mu$. In other words this means that in the ensemble with fixed $\mu$ the magnetic
critical exponents of the system considered belong to the same universality class
as the corresponding lattice model (with the same spatial dimensionality and number
of components of the order parameter and the same properties of the pair
interactions). Of course, the critical temperature $T_c$ of the ferromagnetic
transition (as well as other nonuniversal quantities) will depend on the value of
$\mu$, so that in the ensemble with fixed $\mu$ one has $T_c(\mu)$.

Taking into account this assumption, the thermodynamic potential
near the phase transition point can be written as the sum of two terms
\begin{eqnarray} \label{Rel2}
P = \bar{P}[T^\ast(T,h,\mu), h^\ast(T,h,\mu)] + \Delta P[T,h,\mu] \\
= P_{\rm sing}[T^\ast, h^\ast]  +  P_{\rm reg}[T^\ast, h^\ast]+
\Delta P[T,h,\mu],
\end{eqnarray}
where $\bar{P}[T^\ast, h^\ast]$ describes the critical properties of the `ideal'
system with temperature $T^\ast$ and magnetic field $h^\ast$. $P_{\rm reg}
[T^\ast, h^\ast]$ and $\Delta P [T,h,\mu]$ are nonsingular functions of their
arguments with continuous and smooth derivatives. The scaled temperature
$T^\ast(T,h,\mu)$ and the scaled field $h^\ast(T,h,\mu)$ are also smooth functions,
so that all the singular features of the system in the vicinity of the magnetic
phase transition follows from the properties of $P_{\rm sing}[T^\ast(T,h,\mu),
h^\ast(T,h,\mu)]$. Expression (\ref{Rel2}) can be considered as the mathematical
formulation of the assumption made before. Note that using the ideas of scaling
theory the form of $P_{\rm sing}$ can be further specified. In particular, if the
so-called Wegner corrections \cite{Domb} to asymptotic scaling behaviour are
neglected, one has (see, e.g., \cite{scal,Domb1}) the following expression
\begin{equation} \label{SForm}
P_{\rm sing}[T^\ast, h^\ast]= |\tau^\ast|^{2-\alpha} f(z^\ast),
\end{equation}
where $\tau^\ast=(T^\ast-T_c^\ast)/T_c^\ast$, $z^\ast=h^\ast/|\tau^\ast|^{\beta
+\gamma}$, and $f(z)$ is a universal scaling function.

For the scaled field $h^\ast(T,h,\mu)$, taking into account the symmetry properties
of the magnetic field $h$ we may use the following representation
$$
h^\ast(T,h,\mu) = h \cdot j(T,h,\mu)
$$
with $j(T,h,\mu)= j(T,-h,\mu) >0$. This means that the real transition still
occurs at $h=0$.

For the `ideal' part $\bar{P}[T^\ast, h^\ast]$ we can write down the well
known results for the main thermodynamic quantities. Namely, in the vicinity of
critical point with $h^\ast=0$ and $T^\ast = T_c^\ast$ one has
\begin{equation} \label{Rel4}
m^\ast = \left( \frac{\partial \bar{P}}{\partial h^\ast} \right)_{h^\ast=0} =
A_m |\tau^\ast|^{\beta},
\end{equation}
\begin{equation} \label{Susc}
\chi^\ast = \frac{\partial m^\ast}{\partial h^\ast}
= A_{\chi} |\tau^\ast|^{- \gamma},
\end{equation}
\begin{equation} \label{ent}
s^\ast = \left( \frac{\partial \bar{P}}{\partial T^\ast} \right)_{h^\ast=0} =
s_0 + s_1 \tau^\ast + A_s |\tau^\ast|^{1-\alpha},
\end{equation}
\begin{equation} \label{sp-heat}
c^\ast_v = \frac{\partial s^\ast}{\partial T^\ast}  =
c_0 + c_1 \tau^\ast + A_c |\tau^\ast|^{-\alpha},
\end{equation}
for the magnetization $m^\ast$, the magnetic susceptibility $\chi^\ast$, the
entropy $s^\ast$, and the specific heat $c^\ast_v$ of the `ideal' system,
respectively. The coefficients $A_m$, $A_\chi$, $s_0$, $s_1$, $A_s$, $c_0$,
$c_1$, and $A_c$ in (\ref{Rel4})-(\ref{sp-heat}) do not depend on $\tau^\ast$.
The values of $\beta$, $\gamma$, and $\alpha$ are the corresponding
critical  exponents of the `ideal' system, respectively. It can easily be proved
that the same critical singularities are observed at $h=h^\ast=0$ for the
unconstrained system (\ref{Rel2}) with the fixed  chemical potential, when the
function $T^\ast(T,h,\mu)$ has continuous and smooth derivatives. In this case the
line of critical temperatures $T_c(\mu)$ can be found from the equation
$T^\ast(T_c(\mu),0,\mu)=T_c^\ast$, and using the properties of the function
$T^\ast(T,h,\mu)$ it can be shown that the critical exponents in the grand
canonical ensemble under the assumption made above are the same as for the `ideal'
system.

Let us now consider the critical properties of the constrained system with
the fixed density of particles $n=\bar{n}=$const. In this case, if $h=h^\ast=0$,
one gets from (\ref{Rel2}):
\begin{eqnarray} \label{density}
n  = \left( \frac{\partial P}{\partial \mu} \right)_{T, h=0} =
s^\ast \left(\frac{\partial T^\ast}{\partial \mu}\right)_{T, h=0}
\\ \nonumber + \left( \frac{\partial [P_{\rm reg} + \Delta
P[T,h,\mu]]}{\partial \mu} \right)_{T, h=0}.
\end{eqnarray}
The critical temperature $T_c^0$ in the constrained system can be found from the
condition $T_c^0=T_c (\mu_c)=T_c(\bar{n})$, where the chemical potential $\mu_c$
satisfies the equation $n(\mu_c,T_c (\mu_c))=\bar{n}$ with the function $T_c(\mu)$
defined above. In particular, this gives the equality $T^\ast(T_c^0,0,\mu_c)
\equiv T_c^\ast$, so that in the vicinity of the critical point $[T_c^0,\mu_c, h=0]$,
being of our interest, taking into account the properties of $T^\ast(T,h,\mu)$,
one finds
\begin{equation} \label{Rel3}
T^\ast(T,0,\mu) \simeq T_c^\ast + t_{\mu} \Delta \mu + t_T \Delta T,
\end{equation}
with the coefficients $t_{\mu}$ and $t_{T}$, where $\Delta \mu= \mu -\mu_c$ and
$\Delta T = T-T_c^0$ are assumed to be small enough for restricting of our
consideration by the linear terms in (\ref{Rel3}). This expression can be
rewritten as follows
\begin{equation} \label{Rel3aa}
\tau^\ast \simeq  \bar{t}_{\mu} \Delta \mu + \bar{t}_T \tau
\end{equation}
with $\bar{t}_{\mu}=t_{\mu}/T_c^\ast$ and $\bar{t}_{T}=t_{T}/T_c^\ast$.
In a similar way, taking into account (\ref{ent}), one obtains from
Eq.~(\ref{density}) the following expression for $\Delta n$:
\begin{equation} \label{dens}
\Delta n = n - \bar{n} =  n_1 \tau^\ast + A_n |\tau^\ast|^{1-\alpha} +
n_2 \tau  + n_{\mu} \Delta \mu,
\end{equation}
where $\tau =(T-T_c^0)/T_c^0$. The coefficients $n_1$ and  $A_n$  can be
expressed in terms of the coefficients $s_1$, $A_s$, and $t_{\mu}$, introduced
in (\ref{ent}) and (\ref{Rel3}). For $\Delta n =0$, combining Eqs.~(\ref{Rel3})
and (\ref{dens}), we get the relation between the reduced temperature scales
$\tau^{\ast}$ and $\tau$ in the `ideal' and constrained systems, respectively:

\begin{equation} \label{Main}
\tau = b_0 \tau^{\ast}  + b_1 \tau^{\ast} |\tau^{\ast}|^{-\alpha}
\end{equation}
with
$$
b_0 = \left\{ n_1 +\frac{n_\mu}{\bar{t}_\mu}
\right\} \left(\bar{t}_T \frac{n_\mu}{\bar{t}_{\mu}} - n_2 \right)^{-1}
$$
and
$$
b_1 = A_n \left(\bar{t}_T \frac{n_\mu}{\bar{t}_{\mu}} - n_2 \right)^{-1}.
$$
Similar solutions can be found from (\ref{Rel3aa}) and (\ref{dens}) for
another cases, being of interest in experimental situations. One may be the case
when in the canonical ensemble the temperature $T$ is fixed at the critical value
$T_c^0$, $\tau=0$, but the density $\Delta n$ does change. In this case we obtain
\begin{equation} \label{Main2a}
\Delta n = c_0 \tau^\ast + c_1 \tau^\ast |\tau^\ast|^{-\alpha}
\end{equation}
with $c_0=n_1+n_\mu/\bar{t}_{\mu}$ and $c_1=A_n$. The expressions (\ref{Main})
and (\ref{Main2a}) contain already the central result of Fisher analysis
\cite{Fisher}: it is seen in (\ref{Main}) and (\ref{Main2a}) that, depending on
the sign of the exponent $\alpha$, either the first ($\alpha <0$) or second
($\alpha >0$) term on the right hand side of (\ref{Main}) and/or (\ref{Main2a})
is dominant. This directly leads to Fisher renormalization of the critical
exponents $\beta$ and $\gamma$ when $\alpha$ is positive. In this case the
renormalized exponents $\beta/(1-\alpha)$ and $\gamma/(1-\alpha)$ describe the
critical singularities of magnetization $m$ and magnetic susceptibility $\chi$ in
the constrained system. The critical properties of the specific heat follows
immediately from (\ref{ent}) when we use Eq.~(\ref{Main}) for the scaled reduced
temperature $\tau^\ast$. Taking the derivative in (\ref{ent}) with respect to $T$,
one gets
$$
c_v \sim \frac{\partial s^\ast}{\partial T}  =
\Bigg \{ s_1 + A_s |\tau^\ast|^{-\alpha} \Bigg\}
\frac{1}{T_c^0} \frac{\partial \tau^\ast}{\partial \tau}.
$$
On the other side, if $\alpha >0$, from the expression (\ref{Main}), one has
$$
\frac{\partial \tau^\ast}{\partial \tau} \sim \frac{1}{b_1} |\tau^\ast|^{\alpha}.
$$
Combining these two expressions, it is easily to show that the specific heat
$c_v$ remains finite at the critical point in the constrained system, and a
cusp-like singularity with the exponent $-\alpha/(1-\alpha)$ can be found when
$\alpha >0$. Hence, all the main results of Fisher consideration \cite{Fisher}
are already reproduced for our specific case. Note also that the correction
exponents $\Delta_i$, describing the Wegner corrections to scaling in the `ideal'
system, should be renormalized to $\Delta_i/(1-\alpha)$ if $\alpha$ is positive.
In the opposite case, when $\alpha$ is negative, the critical exponents as well as
the correction exponents $\Delta_i$ in the constrained system remain the same as
in the `ideal' model. These results are known since 1968, when the Fisher's
paper~\cite{Fisher} was published.

\section{Corrections to scaling in the constrained system}

Let us consider some other consequences which follows from the thermodynamic
analysis given in the previous section. Equations (\ref{Main}) and (\ref{Main2a})
may be solved with respect to $\tau^\ast$ by iterations, so that we obtain in
first order
\begin{equation} \label{Main1}
\tau^\ast = {\rm sign}(\tau) \
B |\tau|^{x_{\alpha}} \ \Bigg\{ 1+ b |\tau|^{\Delta_{\alpha}} +
{\cal O}(|\tau|^{2\Delta_{\alpha}})\Bigg\}
\end{equation}
and
\begin{equation} \label{Main2}
\tau^\ast = {\rm sign}(\Delta n) \ C |\Delta n|^{x_{\alpha}} \
\Bigg\{ 1+ c |\Delta n|^{\Delta_{\alpha}} +
{\cal O}(|\Delta n|^{2\Delta_{\alpha}})\Bigg\},
\end{equation}
where the exponents $x_{\alpha}$ and $\Delta_{\alpha}$ depend on the sign of the
critical exponent of specific heat $\alpha$, namely,
\begin{equation} \label{Exp}
x_{\alpha} = \left \{
\begin{array}{ccc}
\displaystyle \frac{1}{1-\alpha}
&, \ \ \alpha > 0,   \\
\ 1  &, \  \  \alpha < 0,
\end{array}
\right.
\end{equation}
and
\begin{equation} \label{Exp1}
\Delta_{\alpha} = \left \{
\begin{array}{ccc}
\displaystyle \frac{\alpha}{1-\alpha} &, \  \ \alpha > 0,   \\
 -\alpha  &,  \  \ \alpha < 0,
\end{array}
\right.
\end{equation}
respectively. The coefficients $B$, $b$ and $C$, $c$ can be expressed by the
initial parameters $b_0$, $b_1$ and $c_0$, $c_1$ in Eqs.~(\ref{Main}) and
(\ref{Main2a}), respectively. For example, one gets the following equations
for the coefficients $B$ and $b$:
\begin{equation} \label{Amp-B}
B = \left \{
\begin{array}{ccc}
\displaystyle b_1^{-1/(1-\alpha)}
&, \ \ \alpha > 0,   \\
\ b_0^{-1}  &, \  \  \alpha < 0,
\end{array}
\right.
\end{equation}
and
\begin{equation} \label{Amp-b}
b = \left \{
\begin{array}{ccc}
\displaystyle - (1-\alpha)^{-1} b_0b_1^{-1/(1-\alpha)}
&, \  \ \alpha > 0,   \\
- b_1 b_0^{-1+\alpha}  &,  \  \ \alpha < 0.
\end{array}
\right.
\end{equation}

It is worth mentioning that the solutions (\ref{Main1}) and (\ref{Main2}) could be
used only under some special demands  which follows form  thermodynamic
stability conditions (see, e.g., \cite{Imry1,Imry2}). In particular, it could
be shown \cite{Imry1} that for $\alpha >0$ a second order phase transition in
the constrained system is observed only if $b_1$ in Eq.~(\ref{Main}) is positive.
For $b_1 <0$ in the vicinity of phase transition (small $\tau^\ast$) the
constrained system becomes unstable and the second order phase transition
transforms to a first order transition. Note that the condition $b_1=0$ gives
the position of a special tricritical point with the critical exponents of the
`ideal' system. Similar consideration may be applied for the opposite case
with $\alpha <0$. Therefore, let us assume that the parameters $b_0$ and $b_1$
($c_0$ and $c_1$) are such that a second order phase transition still exists in
the constrained system, and the main question then arises what are the singular
properties of this transition in the nonasymptotic region.

The first important conclusion can be drawn from Eqs.~(\ref{Main1}) and (\ref{Exp}).
It is seen in Eq.~(\ref{Main1}) that, {\it independently on either the Fisher
renormalization changes the critical exponents in the constrained system or not,
the new corrections to scaling with the exponent $\Delta_{\alpha}$ appear due to
the constraint}. The exponent of these corrections are proportional to $|\alpha|$ and, because this
value is smaller for the Ising, the XY and the Heisenberg universality classes
compared to the Wegner corrections $\Delta_i$, one can expect  significant
contributions from the new corrections to scaling just in the pre-asymptotic
region. For example, the magnetization $m$ in the constrained system when
$\alpha<0$ can be written as follows:
\begin{equation} \label{magnet}
M = A_m |\tau|^{\beta} \Bigg\{1 + a_1 |\tau|^{\Delta_1} +
a_\alpha |\tau|^{|\alpha|} + {\cal O}(|\tau|^{2|\alpha|}) \Bigg\},
\end{equation}
where $\Delta_1=\omega \nu$ is well-known Wegner correction (see, e.g.,
\cite{Domb}). Because of usually $|\alpha| < \Delta_1$, one can conclude, therefore,
that Fisher correction with the exponent $|\alpha|$ is dominant in (\ref{magnet}).
In a similar way the case $\alpha >0$ may be considered. It is evident for both
cases that the width of asymptotic region in temperature scale, where we
can restrict the description of critical singularities by considering the
main critical exponents only, is reduced significantly.

The new corrections to scaling (see (\ref{Main1}) and (\ref{Exp})) with the
exponents (\ref{Exp1}) have appeared in our treatment as the result of the
temperature rescaling due to the constraint. Similar idea was recently used
by Krech \cite{Krech} for the study of spin lattice models with constraints
within the finite size scaling technique. However, we note that the temperature
rescaling is not the only source for corrections of that type. For proving of
this statement let us consider in more details the thermodynamic transformations
from the ensemble with fixed $\mu$ to the ensemble with fixed $n$. But before
proceeding further we derive some additional relations to be useful for the
subsequent calculations.

From Eq.~(\ref{dens}) one derives
\begin{equation} \label{dens-der}
\left(\frac{\partial n}{\partial \mu}\right)_{T,h=0}=
\left[ n_1 + n_c (1-\alpha) |\tau^\ast|^{-\alpha} \right]
\left(\frac{\partial \tau^\ast}{\partial \mu}\right)_{T,h=0} + n_3.
\end{equation}
The derivative $\partial \tau^\ast/\partial \mu$ can be easily found from
Eq.~(\ref{Rel3aa}), and inserting we obtain
\begin{equation} \label{dens-der1}
\left(\frac{\partial n}{\partial \mu}\right)_{T,h=0}=
\left[ c_0 + c_1 (1-\alpha)|\tau^\ast|^{-\alpha} \right] \bar{t}_{\mu},
\end{equation}
where the coefficients $c_0$ and $c_1$ are the same as in Eq.~(\ref{Main2a}).
Recalling now that the derivative (\ref{dens-der}) is directly related to the
compressibility $\kappa_T$ of the system considered ($\left({\partial n}/
{\partial \mu}\right)_{T,h=0}= n^2 \kappa_T$), we can conclude that for $\alpha >0$
a weak divergence appears in the isothermal compressibility in the both ensembles.
The critical exponent describing such singular behaviour in the constrained system
is equal to the renormalized index of specific heat $\alpha/(1-\alpha)$, and
$$
\kappa_T = \kappa_T^0 +  A_\kappa |\tau|^{-\alpha/(1-\alpha)}
$$
with the parameters $\kappa_T^0$ and $A_\kappa = \bar{t}_{\mu} n^2 c_1 (1-\alpha)B$.
This shows the main difference to the case of the specific heat, discussed above.
In the opposite case, when $\alpha$ is negative, the compressibility in the both
ensembles displays the same cusp-like singularity with the exponent $\alpha$ as
is observed in the specific heat behaviour. Hence, our second conclusion is:
{\it for $\alpha>0$ in the constrained system there are some thermodynamic
quantities (e.g., the compressibility in our case) which are asymptotic weak
divergent with the critical exponent $\alpha/(1-\alpha)$}. In general, these
quantities can be expressed by the second order derivatives with respect to a
`hidden' thermodynamic force.

The result obtained above is important also for the study of the corrections
to scaling in thermodynamic derivatives. In order to illustrate this statement
let us use the well-known thermodynamic relation
\begin{equation} \label{Rel5}
{\chi}_{T,n} = \left(\frac{\partial M}{\partial h}\right)_{T,n} =
\left(\frac{\partial M}{\partial h}\right)_{T,\mu} -
\left(\frac{\partial M}{\partial \mu} \right)^2
\left(\frac{\partial n}{\partial \mu} \right)^{-1},
\end{equation}
which gives the connection between the magnetic susceptibility in the canonical
and the grand canonical ensembles. For the first term in the right hand side
of (\ref{Rel5}) the rescaling formula (\ref{Main1}) for temperature dependence
can be used. In the vicinity of the critical point, the contribution from the
second term in (\ref{Rel5}) can be easily estimated  by using Eqs.~(\ref{Rel4}),
(\ref{Rel3aa}), and (\ref{dens-der1}). For the case of positive $\alpha$ this
term produces the contributions proportional to $|\tau^\ast|^{2\beta-2 +\alpha}$,
$|\tau^\ast|^{2\beta-2}$, etc. Applying the hyperscaling relation $2\beta+\gamma=
d\nu$, one obtains for these terms the following estimations $|\tau^\ast|^{-\gamma}$,
$|\tau^\ast|^{-\gamma +\alpha}$, etc., and, taking into account the expression
(\ref{Main1}), one can conclude that an additional correction term with the
exponent $\alpha/(1-\alpha)$ appears in the canonical magnetic susceptibility.
Similar consideration could be used for the case of negative $\alpha$. Hence, the
third conclusion is that {\it in addition to the rescaling mechanism, given by
(\ref{Main1}), there exists another source for the appearance of the same type of
corrections to scaling (proportional to $\alpha$), due to the Legendre
transformation}. In particular, this means that the new corrections to scaling
can not be included in the standard finite size scaling technique just by simple
rescaling of the reduce temperature (\ref{Main1}) as it was proposed in
\cite{Krech}.

The results obtained can be easily generalized for: (i) the case of more complicate
constraint, formulated in the form $F[n,T,\mu]=\theta=$const (see, e.g.,
\cite{Imry1,Imry2}); and (ii) the case, when several `hidden' thermodynamic
forces exist in a system. In the case (i), as it was shown in \cite{Imry1,Imry2},
the generalized constraint will modify the coefficient $b_0$ and $b_1$ in
(\ref{Main}) and change the thermodynamic stability conditions, so that the phase
transition of the second order may transform to a first-order transition. In fact,
the constraint function $F[n,T,\mu]=\theta=$const allows to reformulate all the
results obtained in some `mixed' ensemble with $\theta$ considered as a
thermodynamic parameter, but the conclusions made above will be still valid. Some
exceptions may be noted for special choices of the constraint function $F[n,T,\mu]$
(e.g., in the trivial case $F[n,T,\mu]= f[T,\mu]$).

In case (ii), when several `hidden' thermodynamic forces $\{\mu_i\}$ with
$i=1,2,\ldots, l$ exist in a system and some of them are constrained by external
conditions, we can applied the scheme described above by performing
consecutively Legendre transformations, starting from any constraint
condition. It is evident that, for $\alpha >0$ in the `ideal' unconstrained system,
after the first transformation  we will find renormalized Fisher exponents and
the new corrections to scaling, so that the critical exponent of specific heat
will change sign ($\alpha \rightarrow -\alpha/(1-\alpha)$). This means that the
critical exponents obtained are not changed any in further transformations,
because of negative value of the specific heat exponent $-\alpha/(1-\alpha)$.
Any next Legendre transformation may change only the amplitudes of the
singular terms. Note also that the corrections to scaling, which will appear
after the second (third and so on) Legendre transformation, will have the same
exponents (multiple to $\alpha/(1-\alpha)$) as found after the first step.
The case with $\alpha <0$ in the `ideal' unconstrained system is even more
trivial, and after the first transformation we will find the same (`ideal') critical
exponents with the new corrections to scaling proportional to $|\alpha|$. This
picture will not change in further transformations. Hence, we conclude that {\it
if in the constrained system with arbitrary constraints a second order phase
transition is observed, one has to expect that the critical exponents, describing
this transition in the asymptotic region, are defined by the rules, established by
Fisher \cite{Fisher}, and the leading corrections to scaling are given by
(\ref{Exp1})}.

\section{Discussion and concluding remarks}

We end with a few concluding remarks:

(i) The results presented are quite general and important for many applications.
It is worth to note that constrained equilibrium systems are widely studied in
the literature, but only a few examples could be mentioned when the Fisher
corrections to scaling were included into consideration (see, e.g.,
\cite{Sak,Imry3,Imry4,Hub,MOF}). In general, such corrections may affect
significantly the critical behaviour of compressible magnets
\cite{Sak,Imry3,Imry4}, systems described by Hubbard model \cite{Hub}, binary
and multicomponent mixtures \cite{Anis1,Anis2}, magnetic spin liquids
\cite{Nijmei1,Nijmei2,Ferreira}, and $^3$He-$^4$He mixtures \cite{Ahlers,Folk1}.

(ii) It seems interesting to study more carefully the critical properties of
magnets with  quenched disorder. The motivation is the following. One of
the approaches to the quenched systems is based on the idea of Morita
\cite{Morita,Sobota1,Sobota2}. In this approach a quenched system is considered
as a quasi-equilibrium system with additional forces of constraints, keeping
the quenched system in its equilibrium state and fixing the moments of random
distribution that defines the disordered state. Therefore, the quenched system
can be studied by means of the equilibrium theory, using some results presented
in this paper. In this connection we point out some known results, which support
this view: (a) the so-called Harris criterion \cite{Harris}, proposed for the
quenched systems, is formally similar to the Fisher criterion \cite{Fisher},
concerning the divergence of the specific heat in the constrained system; (b) in
Ref.~\cite{Kyr} within the $\epsilon$-expansion scheme it was proved analytically
(using the replica trick) that the leading correction to scaling in a randomly
diluted inhomogeneous ${\cal O}(m)$ Heisenberg model is a term with the correction
exponent $-\alpha$, as it is expected for the constrained Heisenberg model; (c) in
Ref.~\cite{Dots} it was shown that if $\alpha>0$ in the pure system, the traditional
renormalization group flows, describing the disorder-induced universal critical
behaviour are unstable with respect to replica-symmetry breaking potentials,
found in spin glasses.

(iii) There were several studies of constrained systems, performed within
the renormalization group technique \cite{Sak,Sobota1,Sobota2,Hub,Weg,Rud,Isn1}.
In general the results obtained  support the conclusions made within the
thermodynamic treatment:  Four fixed points (gaussian, spherical, pure model
and renormalized) were found within the renormalization group theory, and
which of these is the stable one (pure or renormalized) depends on the sign of
the exponent $\alpha$ in the `ideal' system. However, we note that, because the
stability condition depends on $\alpha$, it is significant to reconsider
these results in higher approximations and, in particular, to investigate the
effective critical exponents for the most typical renormalization group flows.

(iv) New corrections to scaling, because of smallness $\alpha$ for the Ising,
the XY, and the Heisenberg models, have to be taken into account within standard
finite size scaling technique \cite{MOF}. In particular, we suppose that the
methodological problems found for three- and two-dimensional Ising model fluid
\cite{Nijmei2,Ferreira} are directly connected with the strong crossover effects
caused mainly by Fisher corrections to scaling.

\vspace{4mm}
We thank M.Fisher, M.Anisimov, and Yu.Holovatch for their interest and their
suggestions. Part of this work was supported by the Fonds zur F\"orderung der
wissenschaftlichen Forschung under Project No. P12422-TPH.

\end{multicols}

\end{document}